\documentclass{PoS}

\usepackage{microtype}

\let\OLDthebibliography\thebibliography
\renewcommand\thebibliography[1]{
  \OLDthebibliography{#1}
  \setlength{\parskip}{0pt}
  \setlength{\itemsep}{0pt plus 0.3ex}
}

\title{Simulations of photo-nuclear dijets with Pythia 8 and their sensitivity to nuclear PDFs}

\ShortTitle{Photo-nuclear dijets with Pythia 8}

\author{\speaker{Ilkka Helenius}\\
Institute for Theoretical Physics, T\"{u}bingen University, Auf der Morgenstelle 14, 72076 T\"{u}bingen, Germany\\
E-mail: \email{ilkka.helenius@uni-tuebingen.de}}

\abstract{In ultra-peripheral heavy-ion collisions the nuclei pass with such a large impact parameter that no strong interaction can take place. However, a photon emitted by one nucleus can still interact either with the other nucleus or with a photon from the another nucleus. The former event type corresponds to photoproduction, earlier studied e.g. in electron-proton collisions at HERA, but now with a nuclear target. This provides an unique opportunity to study the nuclear modifications of the PDFs with the photo-nuclear processes measured in lead-lead collisions at the LHC. 
Here we present our recent implementation of photoproduction processes in the \textsc{Pythia~8} general-purpose Monte-Carlo event generator. The resulting simulations are compared to charged-hadron and dijet photoproduction data from the HERA experiments, and the theoretical uncertainties of the framework are studied. We discuss how the relevant part of the photon flux from heavy ions is obtained using the equivalent photon approximation and present predictions for photo-nuclear dijet cross section at the LHC. To estimate the potential of this data to further constrain the nuclear PDFs, the expected statistical uncertainty based on LHC luminosity is compared to uncertainties in the current nuclear PDF analyses.}

\FullConference{XXVI International Workshop on Deep-Inelastic Scattering and Related Subjects (DIS2018)\\
		16-20 April 2018\\
		Kobe, Japan}

\begin{document}

\section{Introduction}


In lepton-hadron collisions the events can be classified in terms of virtuality of the intermediate photon. The high virtuality events ($Q^2\gtrsim 3~\mathrm{GeV}^2$) are referred to as deep inelastic scattering (DIS) events whereas the low virtuality events ($Q^2\lesssim 1~\mathrm{GeV}^2$) are referred to as photoproduction. In the latter case the low virtuality allows factorization of the flux of photons from the hard process cross sections. The hard scale justifying the perturbative treatment needs to be provided by a high-transverse momentum ($p_{\mathrm{T}}$) observable, e.g. jets at sufficiently high $p_{\mathrm{T}}$. In addition to interacting as an unresolved particle, the low virtuality photons may also fluctuate into hadronic states. The partonic structure of these resolved photons can be described with DGLAP-evolved (including a $\gamma \rightarrow q \bar{q}$ source term) photon parton distribution functions (PDFs).

Also high-energy charged hadrons can emit photons that lead to photon-hadron interactions. Due to higher charge the photon luminosity is significantly higher with the heavy ions than with single protons. These interactions can be studied in events where the impact parameter is large enough so that no hadronic interactions take place which would shroud the photon-induced process.
Interest towards these ultra-peripheral collisions (UPCs) has been increasing lately and several new analyses are to be expected soon from different LHC experiments. One main motivation is to use these relatively clean $\gamma A$ processes to constrain nuclear PDFs (nPDFs). 
The virtuality of these UPC photons is very small so they can be treated similarly as photoproduction events in ep collisions.


\textsc{Pythia 8} \cite{Sjostrand:2014zea} is a general purpose Monte-Carlo event generator which have been recently extended for collisions involving leptons and heavy ions. Here the new framework for photoproduction is presented and the results are compared to charged-hadron and dijet photoproduction data in ep collisions at HERA. Furthermore, this framework is applied to calculate dijet cross section in UPCs at the LHC, and the potential of this observable to constrain nuclear PDF is studied.

\section{Framework}

The flux of nearly-collinear photons from charged particles can be obtained by using the equivalent photon approximation (EPA). For leptons the $Q^2$-integrated flux is given by the well-known Weizs\"{a}cker-Williams approximation
\vspace{-0.5em}
\begin{equation}
f_{\gamma}^{\,l}(x_{\gamma}) = \frac{\alpha_{\mathrm{EM}}}
{2\pi} \frac{(1+(1-x_{\gamma})^2)}{x_{\gamma}} \log\left[ \frac{Q^2_{\mathrm{max}}}{Q^2_{\mathrm{min}}(x_{\gamma})} \right],
\label{eq:WW}
\end{equation}
where $x_{\gamma}$ is the momentum fraction of the photon wrt. the momentum of lepton $l$, $Q^2_{\mathrm{min}}$ the minimum virtuality of the photon as limited by the splitting kinematics,
and $Q^2_{\mathrm{max}}$ the maximum allowed virtuality. 
In direct processes the photon act as an unresolved initiator of the hard process and the distribution of photons is simply provided by the flux. In resolved processes the photon flux needs to be convoluted with the photon PDFs to obtain the distribution of partons inside a photon emitted from the beam lepton. When a photon fluctuates into a hadronic state it is also possible to have multiple partonic interactions (MPIs) in the same event. Furthermore, in addition to final-state radiation, also initial-state radiation needs to be simulated for the resolved photons. 

The probability for MPIs is obtained from leading order (LO) cross sections for $2 \rightarrow 2$ QCD processes. These cross sections, however, are divergent in the $p_{\mathrm{T}}\rightarrow 0$ limit and need to be regulated. In \textsc{Pythia 8} the cross sections are made finite by introducing a screening parameter $p_{\mathrm{T 0}}$ which transforms the $1/p_{\mathrm{T}}^4$ divergence into a finite $1/(p_{\mathrm{T 0}}^2 + p_{\mathrm{T}}^2)^2$ behaviour. This is one of the key parameters in \textsc{Pythia 8}, controlling the probability of MPIs and therefore affecting also the yield of small-$p_{\mathrm{T}}$ particles. Therefore one possible way to constrain the energy dependence of this parameter in different collision systems is to compare the $p_{\mathrm{T}}$ spectra to experimental data. 

\begin{figure}[htb]
\begin{minipage}[t]{0.48\textwidth}
\centering
\includegraphics[width=\textwidth]{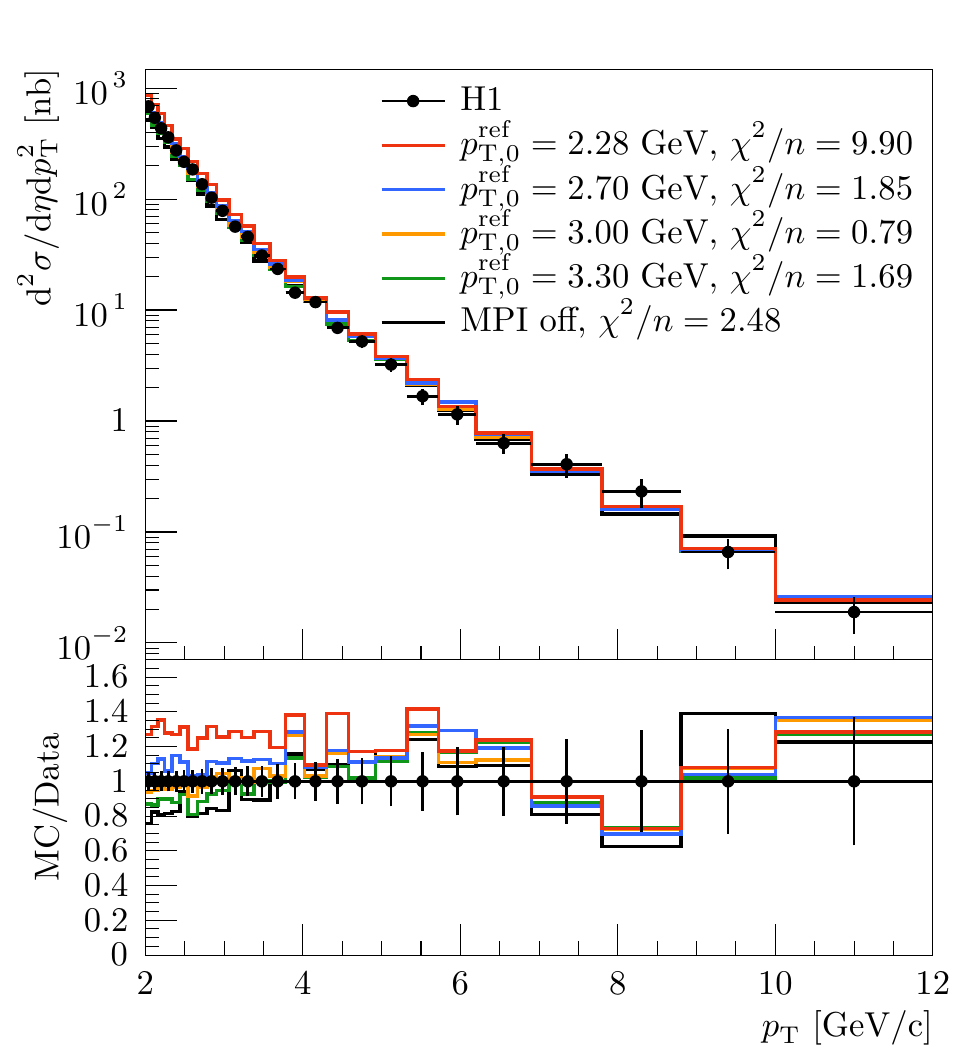}
\caption{Photoproduction of charged hadrons as a function of $p_{\mathrm{T}}$ measured by H1 at HERA. The data is compared to simulations with four different values for $p^{\mathrm{ref}}_{\mathrm{T}0}$ and with MPIs turned off. The resulting $\chi^2/n_{\mathrm{df}}$ are shown for each case.}
\label{fig:H1charged}
\end{minipage}
\begin{minipage}[t]{0.02\textwidth}
\mbox{}
\end{minipage}
\begin{minipage}[t]{0.48\textwidth}
\centering
\includegraphics[width=\textwidth]{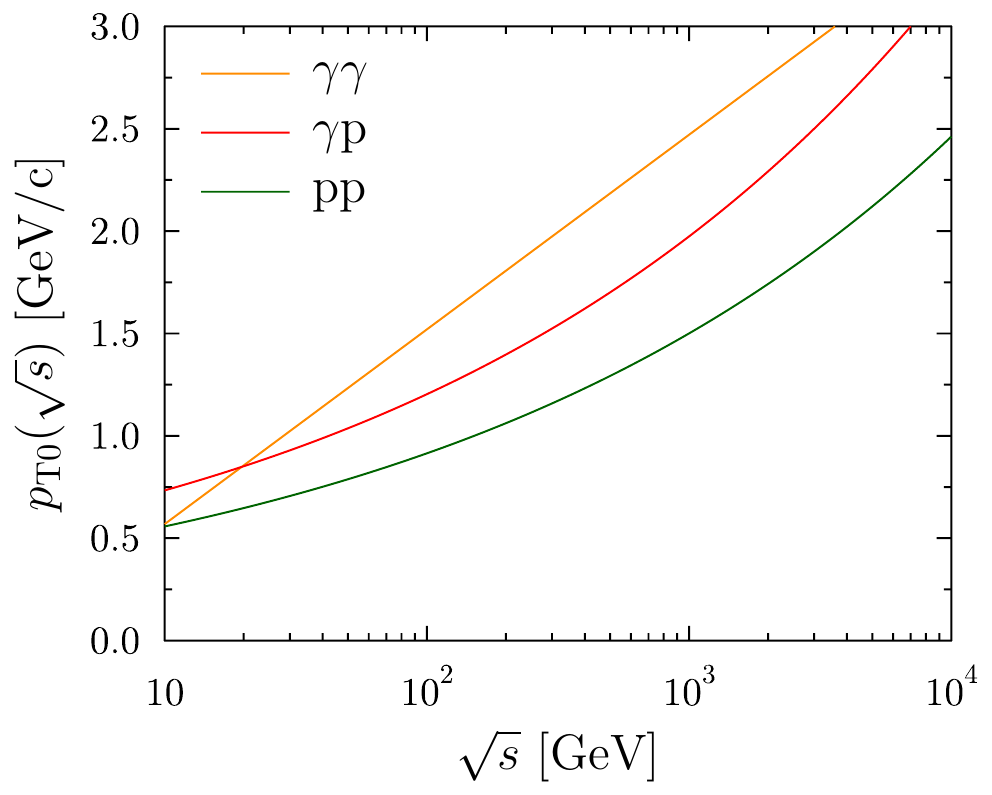}
\caption{Energy dependence of the parameter $p_{\mathrm{T0}}$ for $\gamma$$\gamma$ (orange), $\gamma$p (red) and pp (green) collisions.}
\label{fig:pT0param}
\end{minipage}
\vspace{-1em}
\end{figure}

This kind of comparison is shown in figure \ref{fig:H1charged}, where the generated charged-particle cross section with different values of $p_{\mathrm{T 0}}^{\mathrm{ref}}$ is confronted with H1 measurement \cite{Adloff:1998vt} in ep collisions at HERA. The $p_{\mathrm{T0}}$ is assumed to have a power-law dependence on the collision energy of the photon-proton system, $p_{\mathrm{T}0}(\sqrt{s_{\mathrm{\gamma p}}}) = p^{\mathrm{ref}}_{\mathrm{T}0} (\sqrt{s_{\mathrm{\gamma p}}}/7~\mathrm{TeV})^{\alpha}$, where $\alpha$ has the same value as in default Monash tune for pp collisions, $\alpha = 0.215$. The resulting $\chi^2/n_{\mathrm{df}}$ values are listed in the figure for several choice of $p_{\mathrm{T 0}}^{\mathrm{ref}}$. Using the value from the pp tune ($p^{\mathrm{ref}}_{\mathrm{T}0} = 2.28~\mathrm{GeV/c}$) generates too much low-$p_{\mathrm{T}}$ hadrons whereas the result without any MPIs falls below the data. An optimal description is obtained with $p^{\mathrm{ref}}_{\mathrm{T}0} = 3.0~\mathrm{GeV/c}$ which leads to a $p_{\mathrm{T0}}(\sqrt{s_{\mathrm{\gamma p}}})$ parametrization that sits between the pp tune and a recent parametrization for $\gamma\gamma$ as shown in figure \ref{fig:pT0param}.

\section{Dijet photoproduction at HERA}

The framework can be tested also by comparing to dijet photoproduction in ep collisions measured by different HERA experiments. The advantage is that from the reconstructed jet kinematics one can define observables such as
\begin{equation}
x^{\mathrm{obs}}_{\gamma} = \frac{E^{\mathrm{jet1}}_{\mathrm{T}} \mathrm{e}^{\eta^{\mathrm{jet1}}} + E^{\mathrm{jet2}}_{\mathrm{T}} \mathrm{e}^{\eta^{\mathrm{jet2}}}}{ 2 y E_{\mathrm{e}}},
\label{eq:xgamobs}
\end{equation}
which in an LO parton level calculation would correspond to the momentum fraction of the parton wrt. the photon, and therefore carry some sensitivity whether the jets were produced in a direct or resolved process. Here $E_{\mathrm{T}}^i$ and $\eta^{i}$ refer to the transverse energies and pseudorapidities of the jets, $y$ is the inelasticity of the ep scattering and $E_{\mathrm{e}}$ the energy of the electron beam.

\begin{figure}[htb]
\centering
\includegraphics[width=0.49\textwidth]{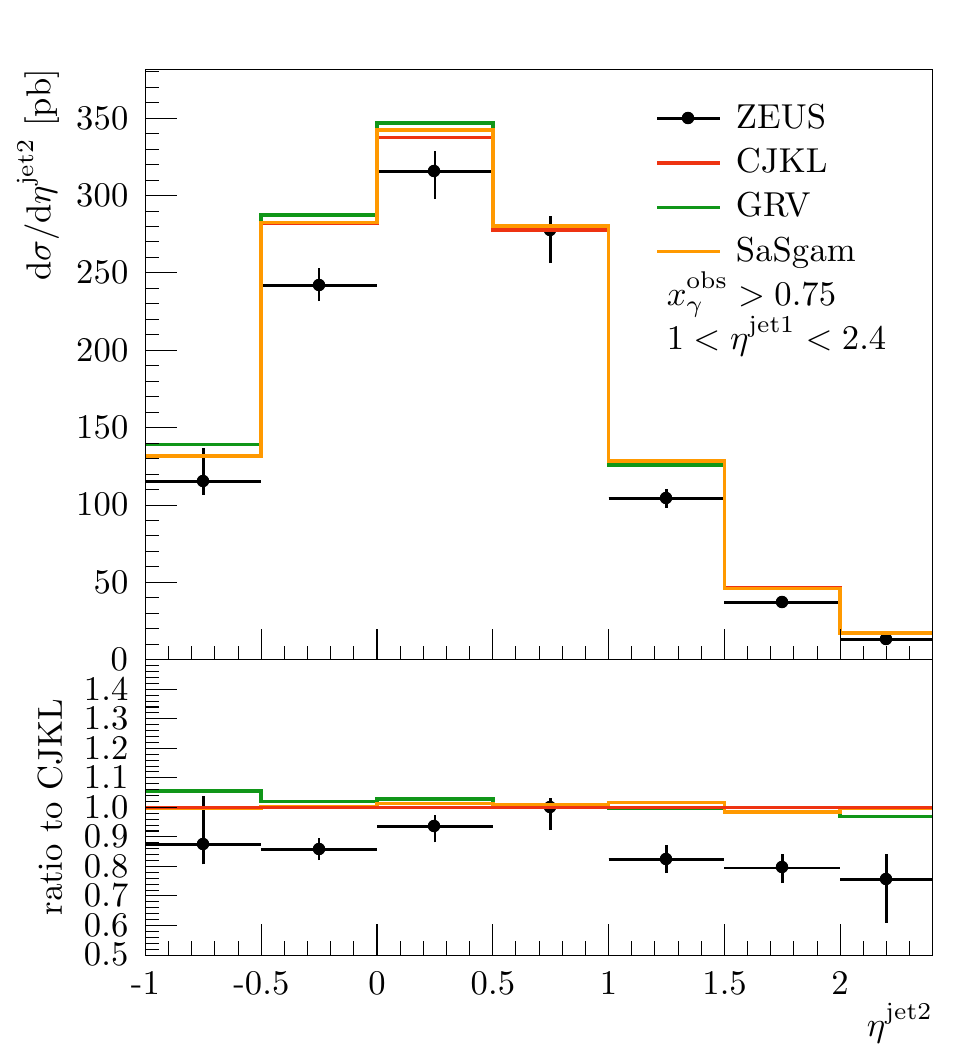}
\includegraphics[width=0.49\textwidth]{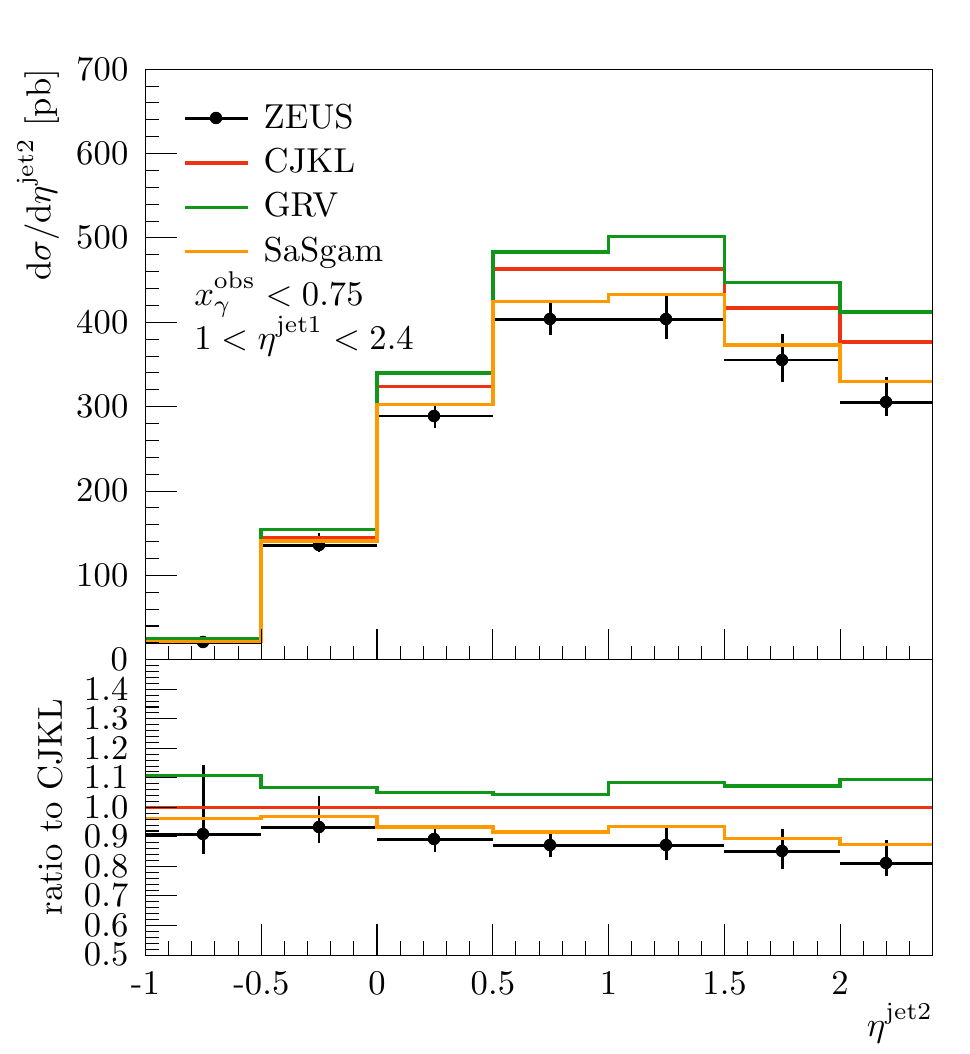}
\label{fig:ZEUSdijet}
\caption{Dijet photoproduction as a function of $\eta^{\mathrm{jet2}}$ for events with $x^{\mathrm{obs}}_{\gamma} < 0.75$ (left) and $x^{\mathrm{obs}}_{\gamma} > 0.75$ (right) with three different photon PDF set, CJKL (red) \cite{Cornet:2002iy}, GRV (green) \cite{Gluck:1991jc} and SaSgam (orange) \cite{Schuler:1995fk}. Data from ZEUS \cite{Chekanov:2001bw}.}
\end{figure}

Figure \ref{fig:ZEUSdijet} show comparisons between the ZEUS measurement \cite{Chekanov:2001bw} for the dijet photoproduction in ep collisions at HERA and the \textsc{Pythia 8} simulation with direct-enhanced ($x^{\mathrm{obs}}_{\gamma} > 0.75$) and resolved-enhanced ($x^{\mathrm{obs}}_{\gamma} < 0.75$) event samples. The (sub-)leading jet were required to have $E_{\mathrm{T}}> 14(11)~\mathrm{GeV}$ and the invariant mass of the photon-proton system were constrained to $134 < W_{\mathrm{\gamma p}}<277~\mathrm{GeV}$ range. Three different set of photon PDFs are used to estimate the theoretical uncertainty arising from the limited precision of the photon PDFs. As expected, for the direct-enriched events the sensitivity to photon PDFs is negligible, whereas for the resolved-enriched this generates $\sim 10~\%$ uncertainty. In general the shape of the dijet cross section is in a good agreement but for both regions of $x^{\mathrm{obs}}_{\gamma}$ the \textsc{Pythia 8} results tend to overshoot the data roughly by $10~\%$.

\section{Ultra-peripheral dijets at the LHC}

\begin{figure}[htb]
\vspace{-1em}
\centering
\includegraphics[width=0.5\textwidth]{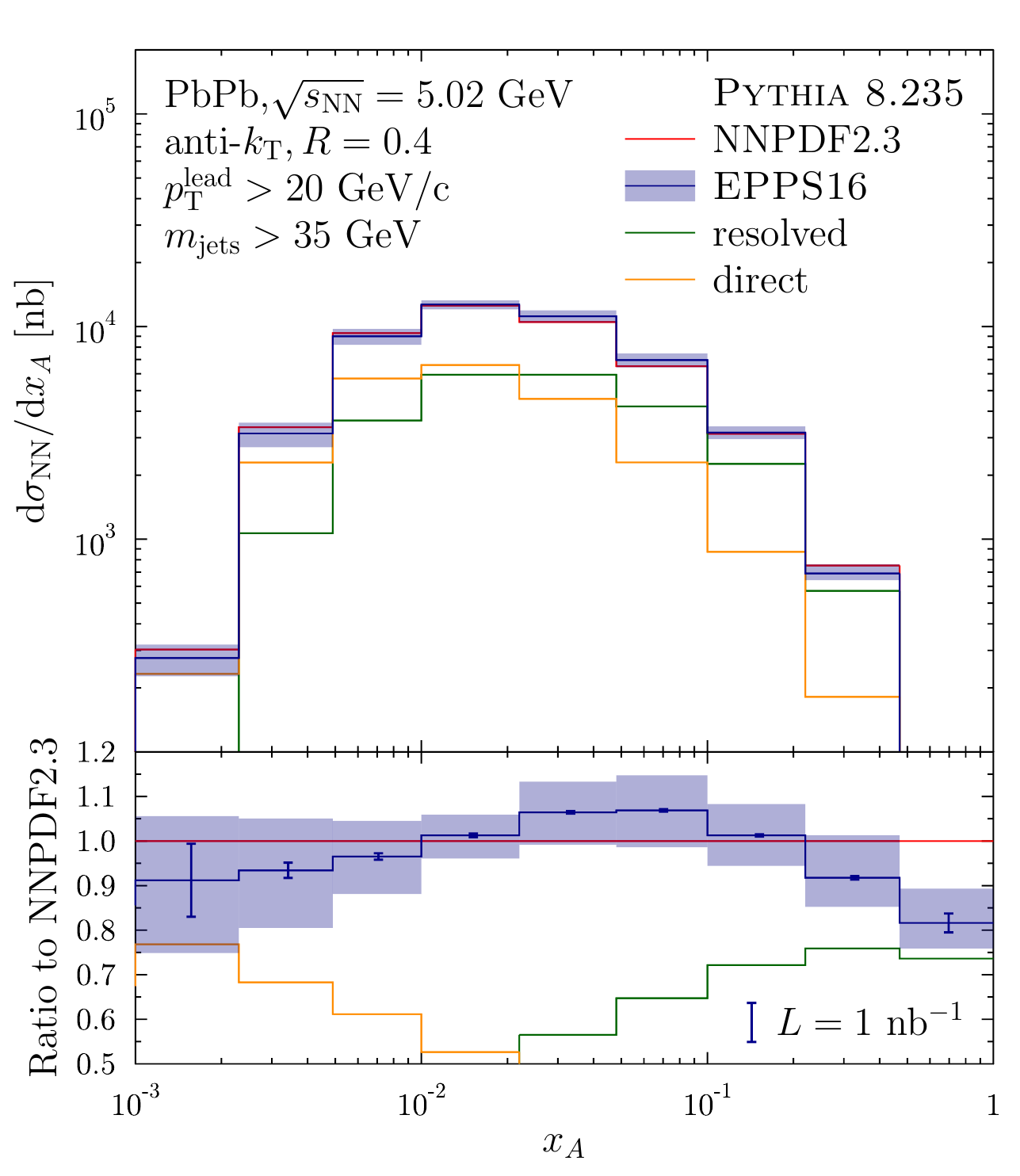}
\vspace{-0.5em}
\caption{Differential per-nucleon cross section for photo-nuclear dijets in ultra-peripheral PbPb collisions at the LHC with ATLAS jet selection. Result using NNPDF2.3 proton PDFs (red) and EPPS16 nuclear modifications (blue) are shown, the uncertainty band is derived from EPPS16 error sets and direct (orange) and resolved (green) contributions are shown separately. The ratio plot shows also expected statistical uncertainties assuming integrated luminosity of $L=1~\mathrm{nb}^{-1}$.}
\label{fig:UPCdijetATLAS}
\end{figure}

The photon flux from heavy-ions can again be obtained from EPA. However, since for ultra-peripheral collisions the events with hadronic interactions need to be rejected, it is more convenient to work in impact-parameter $b$ space instead of momentum space. The $b$-differential form for the flux can be written as
\begin{equation}
f_{\gamma}^{\,A}(x_{\gamma},b) = \frac{\alpha_{\mathrm{EM}} Z^2}{x_{\gamma}  \pi^2} \left[\frac{x_{\gamma}m}{\hbar c} K_1\left(\frac{x_{\gamma}bm}{\hbar c}\right) \right]^2,
\end{equation}
where $Z$ is the charge of the nucleus (number of protons), $m$ the per-nucleon mass and $K_1$ the modified Bessel function of the second kind. The simplest way to perform this rejection is to use a hard-sphere approximation and cut out the events where the colliding nuclei would be in contact.
For photo-nuclear dijets the effective photon flux can then simply be obtained by integrating over the impact parameters larger than the sum of radii of the colliding nuclei $R_A$, yielding 
\begin{equation}
f_{\gamma}^{\,A}(x_{\gamma}) = \frac{\alpha_{\mathrm{EM}} Z^2}{x_{\gamma} \pi} \left[2\xi\, K_1(\xi) K_0(\xi) - \xi^2 \left(K^2_1(\xi) - K^2_0(\xi)  \right) \right],
\end{equation}
where $\xi=2 R_A x_{\gamma} m / (\hbar c)$ and $K_0$ again a modified Bessel function of a second kind. A preliminary ATLAS analysis \cite{ATLAS:2017kwa} defines differential dijet cross sections in terms of three observable: $H_{\mathrm{T}}$, $x_A$ and $z_{\gamma}$ that are derived from the selected jets. Jets are reconstructed using the anti-$k_{\mathrm{T}}$ algorithm with $R=0.4$, and requiring the hardest jet to have $p_{\mathrm{T}}>20~\mathrm{GeV/c}$ and sub-leading jets $p_{\mathrm{T}}>15~\mathrm{GeV/c}$. A particularly interesting observable for nPDF studies is the differential cross section in terms of $x_A$, which is directly related to the probed nuclear-$x$ values. Since a couple of nPDF sets are now included into \textsc{Pythia 8} it is possible to use the presented photoproduction framework to study the potential of this observable to further constrain nPDFs. This differential cross section is shown in figure~\ref{fig:UPCdijetATLAS} using NNPDF2.3 proton PDFs \cite{Ball:2012cx} and with EPPS16 nuclear modifications \cite{Eskola:2016oht}. The error band corresponds to the EPPS16 uncertainty that is compared to the expected statistical uncertainties in the ratio plot assuming an integrated luminosity of $L=1~\mathrm{nb}^{-1}$. In addition, the contribution from direct and resolved processes are shown. At small-$x_A$ region the cross section is dominated by the direct processes, whereas the resolved contribution takes over at $x_A>0.02$. Similarly as for dijet photoproduction at HERA, there is an additional theoretical uncertainty from the photon PDFs for the resolved component. However, this uncertainty is relevant only at larger values of $x_A$ where the resolved contribution dominates.

\section{Summary}

A new framework for photoproduction has been implemented into \textsc{Pythia 8} event generator, including automatic mixing of direct and resolved processes. This has been validated against data from ep collisions at HERA, and the data for charged particle production have been used to constrain the MPI probability in photon-hadron interactions. The framework has been applied to simulate ultra-peripheral heavy ion collisions, where the photon flux is provided by the high-energy ions. These processes allow to study rather clean photo-nuclear interactions that can be used to study nuclear structure. Here we have demonstrated that the photo-nuclear dijets at the LHC can provide new constraint for nPDFs at $x_A > 10^{-3}$ when using similar jet selection as in the preliminary ATLAS study \cite{ATLAS:2017kwa}.

\acknowledgments
\noindent This work have been supported by Carl Zeiss Foundation and has received funding from the European Research Council (ERC) under the European Union's Horizon 2020 research and innovation programme (grant agreement No 668679).

\bibliographystyle{JHEP}
\bibliography{HeleniusDIS2018}

\providecommand{\href}[2]{#2}\begingroup\raggedright\begin{thebibliography}{10}

\bibitem{Sjostrand:2014zea}
T.~Sj{\"o}strand, S.~Ask, J.~R. Christiansen, R.~Corke, N.~Desai, P.~Ilten
  et~al., \emph{{An Introduction to PYTHIA 8.2}},
  \href{http://dx.doi.org/10.1016/j.cpc.2015.01.024}{\emph{Comput. Phys.
  Commun.} {\bf 191} (2015) 159--177},
  [\href{http://arxiv.org/abs/arXiv:1410.3012}{{\tt arXiv:1410.3012}}].

\bibitem{Adloff:1998vt}
{\scshape H1} collaboration, C.~Adloff et~al., \emph{{Charged particle
  cross-sections in photoproduction and extraction of the gluon density in the
  photon}}, \href{http://dx.doi.org/10.1007/s100520050761}{\emph{Eur. Phys. J.}
  {\bf C10} (1999) 363--372}, [\href{http://arxiv.org/abs/hep-ex/9810020}{{\tt
  hep-ex/9810020}}].

\bibitem{Chekanov:2001bw}
{\scshape ZEUS} collaboration, S.~Chekanov et~al., \emph{{Dijet photoproduction
  at HERA and the structure of the photon}},
  \href{http://dx.doi.org/10.1007/s100520200936}{\emph{Eur. Phys. J.} {\bf C23}
  (2002) 615--631}, [\href{http://arxiv.org/abs/hep-ex/0112029}{{\tt
  hep-ex/0112029}}].

\bibitem{Cornet:2002iy}
F.~Cornet, P.~Jankowski, M.~Krawczyk and A.~Lorca, \emph{{A New five flavor LO
  analysis and parametrization of parton distributions in the real photon}},
  \href{http://dx.doi.org/10.1103/PhysRevD.68.014010}{\emph{Phys. Rev.} {\bf
  D68} (2003) 014010}, [\href{http://arxiv.org/abs/hep-ph/0212160}{{\tt
  hep-ph/0212160}}].

\bibitem{Gluck:1991jc}
M.~Gluck, E.~Reya and A.~Vogt, \emph{{Photonic parton distributions}},
  \href{http://dx.doi.org/10.1103/PhysRevD.46.1973}{\emph{Phys. Rev.} {\bf D46}
  (1992) 1973--1979}.

\bibitem{Schuler:1995fk}
G.~A. Schuler and T.~Sjostrand, \emph{{Low and high mass components of the
  photon distribution functions}},
  \href{http://dx.doi.org/10.1007/BF01565260}{\emph{Z. Phys.} {\bf C68} (1995)
  607--624}, [\href{http://arxiv.org/abs/hep-ph/9503384}{{\tt
  hep-ph/9503384}}].

\bibitem{ATLAS:2016vdy}
{\scshape ATLAS} collaboration, \emph{{Measurement of high-mass dimuon pairs
  from ultraperipheral lead-lead collisions at $\sqrt{s_{\mathrm{NN}}}=5.02$
  TeV with the ATLAS detector at the LHC}}, {\emph{ATLAS-CONF-2016-025} (2016)}.

\bibitem{Klein:2016yzr}
S.~R. Klein, J.~Nystrand, J.~Seger, Y.~Gorbunov and J.~Butterworth,
  \emph{{STARlight: A Monte Carlo simulation program for ultra-peripheral
  collisions of relativistic ions}},
  \href{http://dx.doi.org/10.1016/j.cpc.2016.10.016}{\emph{Comput. Phys.
  Commun.} {\bf 212} (2017) 258--268},
  [\href{http://arxiv.org/abs/arXiv:1607.03838}{{\tt arXiv:1607.03838}}].

\bibitem{ATLAS:2017kwa}
{\scshape ATLAS} collaboration, \emph{{Photo-nuclear dijet production in
  ultra-peripheral Pb+Pb collisions}}, {\emph{ATLAS-CONF-2017-011} (2017)}.

\bibitem{Ball:2012cx}
R.~D. Ball et~al., \emph{{Parton distributions with LHC data}},
  \href{http://dx.doi.org/10.1016/j.nuclphysb.2012.10.003}{\emph{Nucl. Phys.}
  {\bf B867} (2013) 244--289}, [\href{http://arxiv.org/abs/1207.1303}{{\tt
  1207.1303}}].

\bibitem{Eskola:2016oht}
K.~J. Eskola, P.~Paakkinen, H.~Paukkunen and C.~A. Salgado, \emph{{EPPS16:
  Nuclear parton distributions with LHC data}},
  \href{http://dx.doi.org/10.1140/epjc/s10052-017-4725-9}{\emph{Eur. Phys. J.}
  {\bf C77} (2017) 163}, [\href{http://arxiv.org/abs/arXiv:1612.05741}{{\tt
  arXiv:1612.05741}}].

\end{thebibliography}\endgroup

\end{document}